# Magnetic states and ferromagnetic resonance in geometrically frustrated arrays of multilayer ferromagnetic nanoparticles ordered on triangular lattices

V.L. Mironov[1,2*], E.V. Skorohodov[1], J.A. Blackman[3]

[1] Department of magnetic nanostructures, Institute for physics of microstructures RAS, Nizhniy Novgorod, GSP-105, 603950, Russia

[2] Department of physics and nanoelectronics, Lobachevsky State University of Nizhniy Novgorod, Nizhniy Novgorod, Gagarin Avenue, 23, 603950, Russia

[3] Department of Physics, University of Leicester, Leicester, LE1 7RH, UK

We present a theoretical investigation of magnetostatic interaction effects in geometrically frustrated arrays of anisotropic multilayer ferromagnetic nanoparticles arranged in different spatially configured systems with triangular symmetry. We show that the interlayer magnetostatic interaction significantly expands the opportunities to create magnetically frustrated systems. The effects of the magnetostatic interaction in magnetization reversal processes and the possibility to control the ferromagnetic resonance spectrum in such systems are discussed.

## 1. Introduction

The effects of magnetostatic interaction in the regular arrays of anisotropic single-domain ferromagnetic nanoparticles arranged in two-dimensional lattices with different spatial symmetry (in the literature, such systems are called "artificial spin ice") are the subject of the intensive investigations in the last decade [1-6]. The increasing interest in these objects is associated primarily with the possibility to investigate the fundamental properties of geometrically frustrated magnetic systems using a relatively simple model.

The recent progress in the e-beam nanolithography techniques enables the fabrication of superdense arrays of nanoparticles, which demonstrate an unusual collective behavior connected with the competition between configuration entropy effects, dipolar interaction and artificial anisotropy [7-9]. It was shown that artificial spin ice systems demonstrate the considerable changes in coercivity, switching fields and scenario of magnetization reversal. In particular, the dynamic switching in an external magnetic field is accompanied by the effects of the effective magnetic charges ordering [9] and the appearance of the exotic states called "magnetic monopoles" [10,11].

The basic idea of our work is the investigation of the influence of magnetostatic interaction on the magnetic states and high frequency properties of the frustrated magnetic systems based on multilayer stacks of binary nanoparticles. In this case one can expect a considerable expansion of the spectrum of magnetic states and a significant increase in the averaged magnetostatic energy at a lattice site due to the strong dipolar coupling between particles in the neighboring layers. Furthermore, in multilayer systems the effects of geometrical frustration can be significantly expanded due to the variation of the magnetic moments for the particles located in different layers by varying the materials and layer thicknesses.

In the current letter we concentrate our attention on two aspects of the magnetostatic interaction in multilayer stack arrays. First is the realization of exotic magnetic configurations in the magnetization reversal process, which remain stable after removal of the external magnetic field. The other interesting problem is the influence of intralayer and interlayer magnetostatic interactions on the spectrum of ferromagnetic resonance and magnetostatic spin waves in the ordered arrays of multilayer stacks as the prototypes of artificial 3-D magnonic crystals [12-15]. In particular, we show that the spectrum of collective modes of ferromagnetic resonance in such systems strongly depends on the spatial configuration of the magnetic moments of particles and can be significantly changed by switching of magnetic states in an external magnetic field. From a practical point of view such systems are promising for the development of tunable microwave devices [16-18] for civil and military applications.

## 2. The methods of calculations

We investigated the effects of magnetostatic interaction and ferromagnetic resonance in arrays of the elliptical nanodisks ($a \times b \times h$) arranged in a triangular grating (Fig. 1). To simplify the calculations we used the theoretical model of anisotropic dipoles commonly used for the description of such systems [4, 19, 20]. We assumed that the magnetic field of the nanoparticles corresponds to the field of a uniformly magnetized sphere with built-in anisotropy corresponding to the shape anisotropy of an elliptical disc. In this approximation the energy of a system can be presented as

$$W = \frac{1}{2}\sum_{i \neq j}\left\{\frac{(\vec{M}_i \cdot \vec{M}_j)}{R_{ij}^3} - \frac{3(\vec{M}_i \cdot \vec{R}_{ij})(\vec{M}_j \cdot \vec{R}_{ij})}{R_{ij}^5}\right\} + \frac{1}{2}\sum_i\left\{N_{xi}M_{xi}^2 + N_{yi}M_{yi}^2 + N_{zi}M_{zi}^2\right\} - \sum_i\left(\vec{M}_i \cdot \vec{H}_{ex}\right), \qquad (1)$$

---

[*] Corresponding author: mironov@ipmras.ru

where $\vec{M}_i$ is magnetic moment of *i*-th particle; $N_{xi}$, $N_{yi}$ and $N_{zi}$ are demagnetizing factors along the main axes of elliptical disc; $R_{ij}$ is the separation between disc's centers; $\vec{H}_{ex}$ is an external magnetic field.

To find the eigenfrequencies of the ferromagnetic resonance in the particle arrays we solved the system of Landau-Lifshitz equations without external magnetic field. The oscillations of the magnetic moment of *i*-th particle in the array are described by the following equation:

$$\frac{\partial \vec{M}_i}{\partial t} = -\gamma \left[ \vec{M}_i \times \vec{H}_i \right], \qquad (2)$$

where $\gamma$ is gyromagnetic ratio, $\vec{H}_i$ is the effective magnetic field, which is generally defined as

$$\vec{H}_i = \sum_{j \neq i} \vec{H}_j - \vec{H}_{di}. \qquad (3)$$

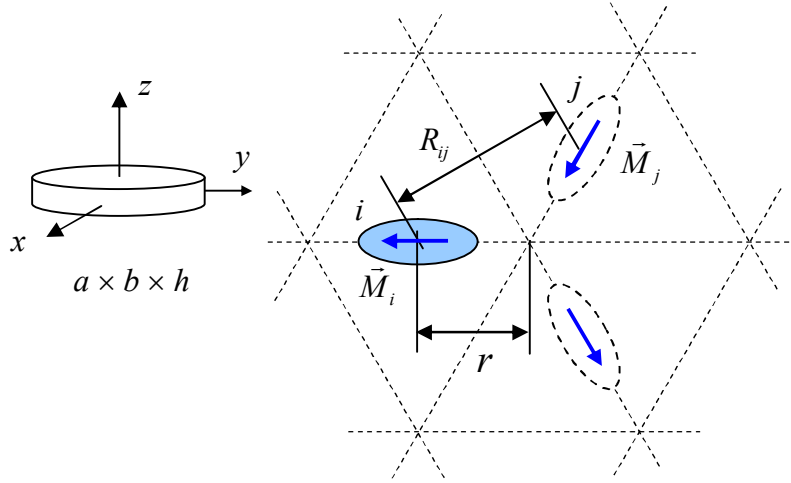

**Fig. 1.** The schematic drawing of the elliptical particles array on triangular lattice.

Here $\vec{H}_j$ is the stray field from *j*-th particle at the location of *i*-th particle and $\vec{H}_{di}$ is demagnetizing field, which is defined as

$$\vec{H}_{di} = -\vec{\vec{N}} \vec{M}_i,$$

where $\vec{\vec{N}}$ is the tensor of demagnetizing factors:

$$\vec{\vec{N}} = \begin{pmatrix} N_x & 0 & 0 \\ 0 & N_y & 0 \\ 0 & 0 & N_z \end{pmatrix}.$$

The FMR eigenfrequency of the single elliptical disc is defined by effective fields of shape anisotropy

$$\omega_0 = \gamma \sqrt{(N_x - N_y)(N_z - N_y)} M_S, \qquad (4)$$

where $M_S$ is the saturation magnetization.

To find the spectrum of eigenfrequencies for the interacting particles array we solved the linearized equation (2) with $\vec{H}_{ex} = 0$. The magnetic moment of each particle and acting field were represented as:

$$\vec{M}_i = \vec{M}_i^{st} + \vec{m}_i(t),$$
$$\vec{H}_j = \vec{H}_j^{st} + \vec{h}_j,$$

where $\vec{M}_i^{st}$ is an average static magnetic moment of the particle; $\vec{m}_i(t)$ is an alternating magnetic moment $\left( \vec{m}_i(t) \ll \vec{M}_i^{st} \right)$; $\vec{H}_j^{st}$ is a static field; $\vec{h}_j$ is a high frequency magnetic field acting on the *i*-th particle from the j-th particle, ( $\vec{h}_j \ll \vec{H}_j^{st}$ ). Then the linearized system of Landau-Lifshitz equations can be written as follows:

$$\frac{\partial \vec{m}_i}{\partial t} = -\gamma \left[ \vec{M}_i^{st} \times \left( \sum_{j \neq i} \vec{h}_j - \hat{N}\vec{m}_i \right) \right] - \gamma \left[ \vec{m}_i \times \left( \sum_{j \neq i} \vec{H}_j^{st} - \hat{N}\vec{M}_i^{st} \right) \right]. \quad (5)$$

Further, we always assume that the effective field associated with the particle shape anisotropy more than the fields of the other particles ($(N_x - N_y)M_i$; $(N_z - N_y)M_i \gg \vec{H}_j^{st}$), i.e. precession of the magnetic moments occurs mainly around the anisotropy axes of each particle. However the spectrum of eigenfrequencies in such systems is determined by the magnetostatic interaction and strongly depends on the spatial configuration of the magnetic moments.

The remagnetization effects in interacting arrays under an external magnetic field were investigated by Monte Carlo simulations based on the model of anisotropic spheres. In the calculations we have neglected any deviation of the vectors $\vec{M}_i$ out of the plane of the elliptical disc and used the simple model of uniaxial anisotropy. In this case the magnetic energy of particles array can be represented as

$$W = \frac{1}{2} \sum_{i \neq j} \left\{ \frac{(\vec{M}_i \cdot \vec{M}_j)}{R_{ij}^3} - \frac{3(\vec{M}_i \cdot \vec{R}_{ij})(\vec{M}_j \cdot \vec{R}_{ij})}{R_{ij}^5} \right\} + KV \sum_i \sin^2(\theta_i - \theta_i^*) - \sum_i \left( \vec{M}_i \cdot \vec{H}_{ex} \right), \quad (6)$$

where $K$ is anisotropy constant; $V$ is the volume of particle; $\theta_i$ is the angle characterizing the direction of magnetic moment of $i$-th particle; $\theta_i^*$ is the angle characterizing the direction of anisotropy axis of $i$-th particle. The normalized energy can be represented in the following form:

$$W = \alpha \varepsilon + \sum_i \sin^2(\theta_i - \theta_i^*) - \sum_i \left( \vec{m}_i \cdot \vec{H}_{ex}^* \right). \quad (7)$$

Here $\varepsilon$ is the normalized magnetostatic energy

$$\varepsilon = \frac{1}{2} \sum_{i \neq j} \left\{ \frac{(\vec{m}_{0i} \cdot \vec{m}_{0j})}{r_{ij}'^3} - \frac{3(\vec{m}_{0i} \cdot \vec{r}_{ij}')(\vec{m}_{0j} \cdot \vec{r}_{ij}')}{r_{ij}'^5} \right\}, \quad (8)$$

where $\vec{m}_{0i}$ is a unit vector of the magnetic moment of $i$-th particle, $r_{ij}'$ is normalized separation between particle centers. The parameter $\alpha$ and normalized external magnetic field $\vec{H}_{ex}^*$ are expressed as follows:

$$\alpha = \frac{M_s^2 V}{R^3 K}, \quad (9)$$

$$H_{ex}^* = \frac{H_{ex}}{H_{an}}, \quad (10)$$

where $H_{an}$ is effective field of anisotropy ($H_{an} = K/M_s$); $M_s$ is a saturation magnetization of particle material. The parameter $\alpha$ is the ratio of the magnetostatic energy to the energy of anisotropy.

The magnetization reversal processes and stationary distribution of the magnetic moments in the particle's arrays corresponding to the minimum energy (7) were simulated by the Monte Carlo method.

## 3. Magnetostatic interaction in one-layer systems

One of the simplest systems of interacting binary particles is a system with basic a building block of three particles located in the plane at an angle of 120° (Fig. 2).

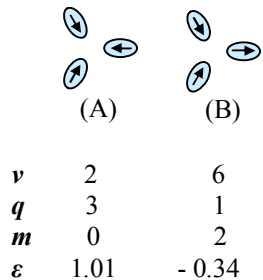

|   | (A)  | (B)   |
|---|------|-------|
| *v* | 2    | 6     |
| *q* | 3    | 1     |
| *m* | 0    | 2     |
| *ε* | 1.01 | − 0.34 |

**Fig. 2.** The possible configurations of magnetic moments in an array of three particles with triangular symmetry. Parameters: *v* is the number of ways to implement the state; *q* is an effective magnetic charge on the lattice site; *m* is an effective magnetic moment on the lattice site; *ε* is normalized density of magnetostatic energy.

As it is known this system has two states differing in the spatial configuration of the magnetic moments [6]. These states differ in the probability of their realization at random initial conditions (this quantity is proportional to the number of ways to implement this state *v*), the effective magnetic charge on the lattice site *q*, the effective magnetic moment on the lattice site *m*, and the normalized density of magnetostatic energy *ε*. As is seen, the state (A) corresponding to the highest effective magnetic charge and zero mean magnetic moment ("magnetic monopole") has the highest energy. The quasi-homogeneous state (B) with effective charge *q* = 1 and nonzero average magnetic moment has a lower energy.

*3.1. Magnetization reversal in the one-layer three particles array*

The form of the magnetization reversal of the three particle array in an uniform external magnetic field depends on the ratio, $\alpha$, of the anisotropy energy to the magnetostatic interaction. The hysteresis curves and schemes of transitions between different states for the array of three interacting particles are shown in Fig. 3.

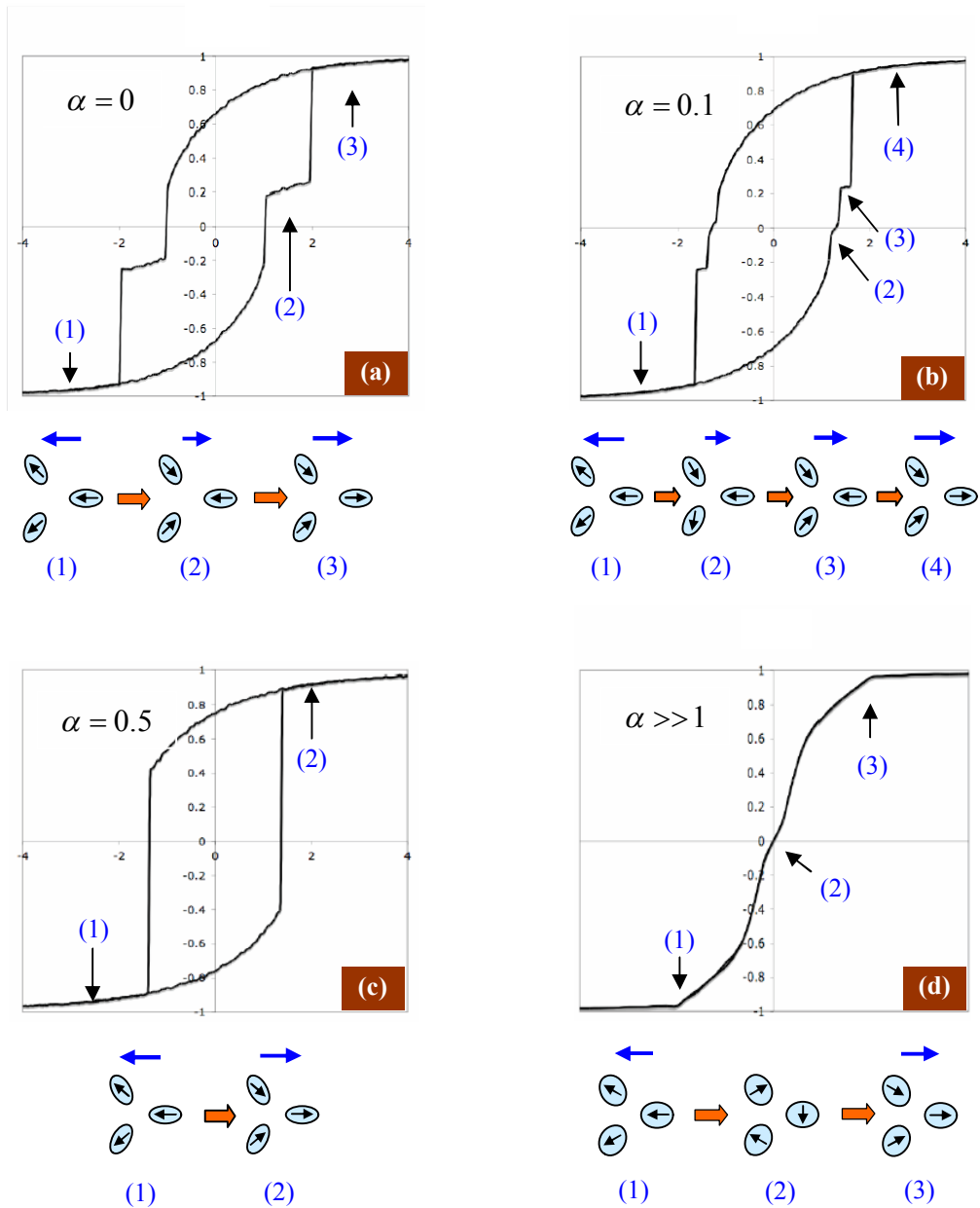

**Fig. 3.** The hysteresis curves and the schemes of transitions between different states of three particles array: (a) $\alpha = 0$, (b) $\alpha = 0.1$, (c) $\alpha = 0.5$, (d) $\alpha \gg 1$. The configurations of the magnetic moments of the particles corresponds to the points on the hysteresis curves. The direction of the external magnetic field is shown in the diagrams by arrows located above the array of particles.

If the anisotropy is dominant as in Fig. 3a (here, the magnetostatic interaction is switched off completely, α=0), the high energy state is realized ("magnetic monopole" state (2) in Fig. 3a) at intermediate reversing field. In this case we can prepare two stable states (A) and (B), in the notation of Fig. 2, by an external magnetic field. As the interaction energy is increased in comparison with anisotropy energy, the high energy state (B) becomes unstable. For example, the hysteresis curve for $\alpha = 0.1$ (in Fig. 3b) demonstrates only a narrow step corresponding to the state (B), while for α=0.5 (Fig 3c) there is a continuous transition between state (A) and its complete reversal. As the interaction energy is further increased (negligible anisotropy), $\alpha \gg 1$, as in Fig. 3d, the hysteresis loop collapses. The ground state of the array at zero external magnetic field is the vortex state (state (2) in the transition scheme of Fig. 3d).

### 3.2. Ferromagnetic resonance in single layer systems

We calculated the dependence of the the FMR frequencies of the three particle array on parameter $r$ (the distance between center of particle and center of array). The increasing magnetostatic interaction between particles leads to a splitting of the FMR spectrum. The dependences of normalized FMR frequencies on the nondimensional parameter $V/r^3$ for the region of parameter $\alpha < 0.1$ are shown in Fig. 4.

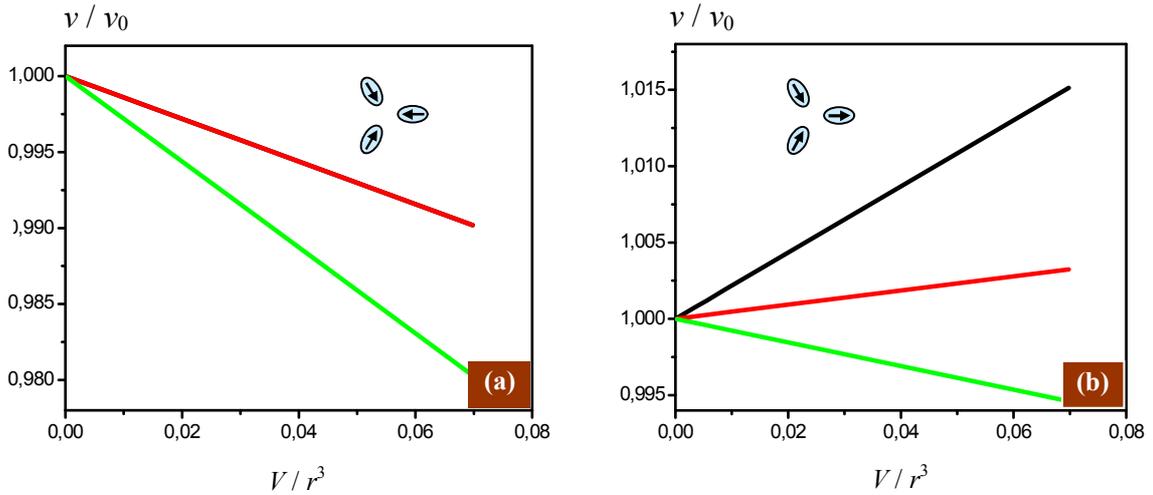

**Fig. 4.** The dependences of normalized FMR frequencies on the distance between center of the particles and array center for two possible magnetic moment configurations. The FMR frequencies are normalized on the resonance frequency of single particle.

As is seen from Fig. 4a the interaction between particles in a high-energy configuration (A) leads to the splitting of the FMR spectrum into two branches and shifting to lower frequencies. On the other hand, for the low-energy configuration (B) the splitting into three branches is observed (Fig 4b). Two branches are shifted to the higher frequencies and one to lower frequencies. Since the high-energy configuration (A) has a higher symmetry than configuration (B) the lower branch of spectrum in Fig. 4a is doubly degenerate.

The three elements group considered above can be combined into the structures with different symmetry, leading to additional changes in the spectrum of FMR. Here we consider three possible spatial configurations of association in symmetrical two-dimensional arrays (Fig. 5) and discuss the changes in the FMR spectra from parameter $r^*$ (distance between centers of particle groups).

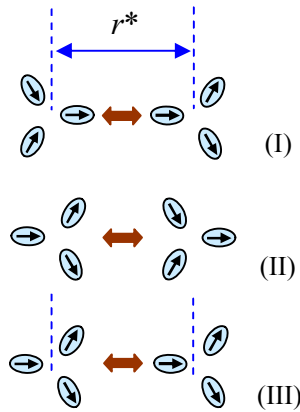

**Fig. 5.** Three spatial configurations with quasi-homogeneous state for association of three particle group. The parameter $r^*$ is the distance between canters of particle groups.

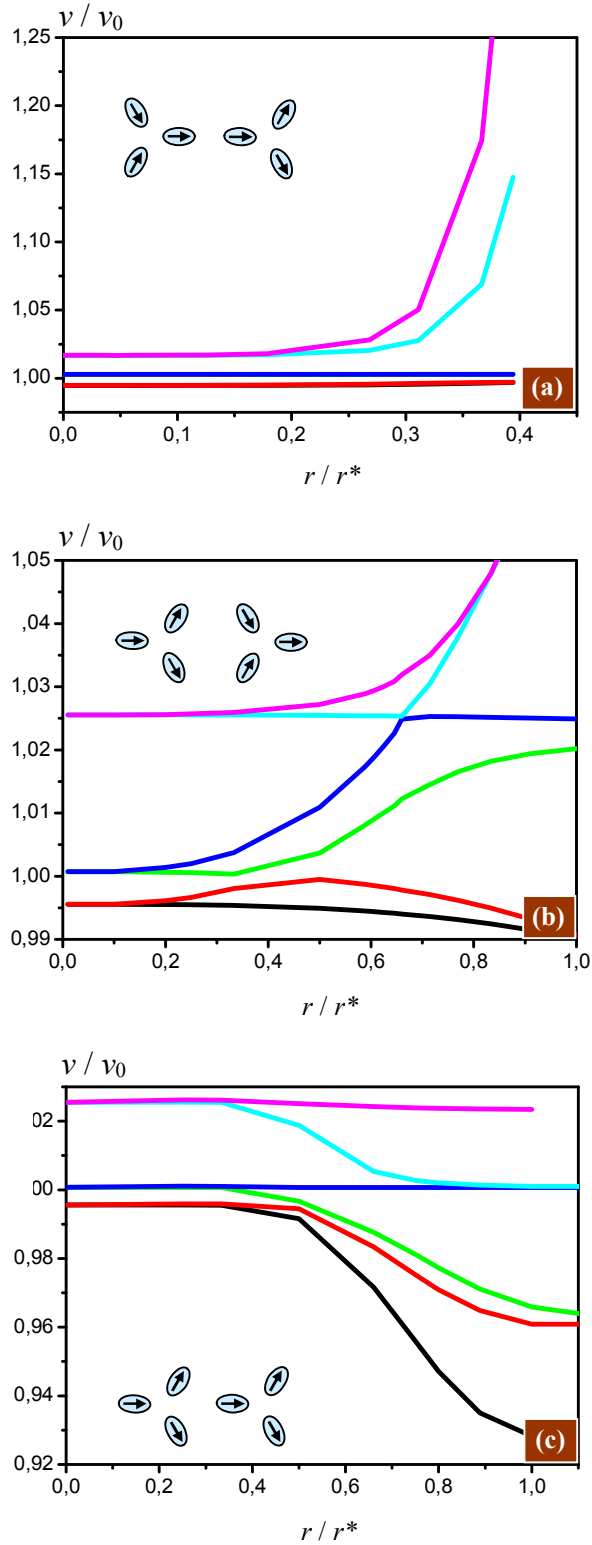

**Fig. 6.** The dependences of normalized FMR frequencies on the distance between centers of particle groups.

In the configuration (I) arrays interact mainly only by the two outermost particles. This leads to an additional monotonic splitting of the high frequency modes in the FMR spectrum (see Fig. 6a and compare with Fig. 5b). A different situation is observed for the configurations (II) and (III). The interaction energy of the arrays in these configurations is less than in the configuration (I), and as a result lower frequency splitting is observed (see Fig. 6b and Fig. 6c). However these configurations implemented a nonmonotonic changing FMR spectrum that is the effect of competition between interaction of three elements inside group and intergroup interaction.

## 4. The two-layer systems

The magnetostatic interaction energy strongly depends on the distance between the particles and can be estimated as

$$W_{int} \approx \frac{M^2 V^2}{d^3}.$$

In this connection the stack arrangement of elliptical particles (one above the other, see Fig. 7a) leads to an essential increase in the interaction energy. In addition, it can be easy realized technologically.

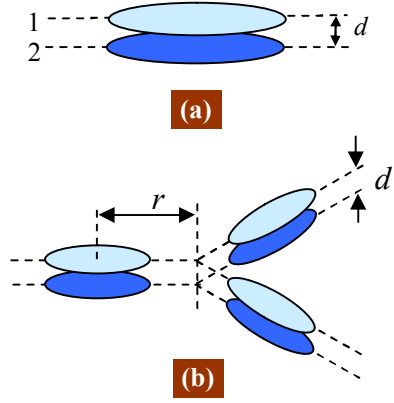

**Fig. 7.** The schematic image of two-layer stack (a) and array of three two-layer particles with triangular symmetry (b).

We investigated the peculiarities of the magnetostatic interaction in an array of two-layer particles with triangular symmetry (Fig. 7b). The dependences of magnetostatic interaction energies for different configurations of magnetic moments on the interlayer separation are presented in Fig. 8 (a). When the distance between the layers is large (parameter $r/d$ is close to zero) the interaction between the layers is small and in this system three energy-degenerate states (I), (II) and (III) are implemented. The degrees of degeneracy are (I) – 2; (II) – 2; (III) – 4 respectively (see Fig. 8b). The close approach of layers removes the degeneracy and the energy spectrum is split into eight components due to the interlayer interaction. The energies of one group states are shifted to higher energies (u) and the energies of other states are shifted to lower energies (d). In the region of parameters $r/d$ $0.6 \div 1$ we observe the crossing of energy levels from completely different configurations (I, II and III) as the interaction of particles in the layer becomes comparable with the interaction of the particles located in different layers.

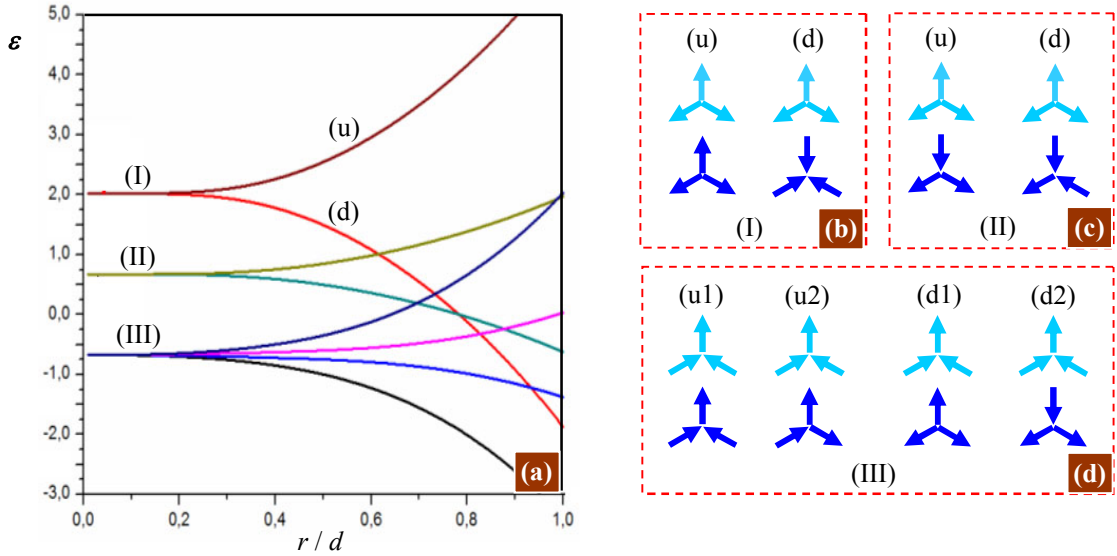

**Fig. 8.** (a) is the dependences of normalized magnetostatic energies for different configurations of magnetic moments on interlayer separation. (b), (c) and (d) are possible configurations of magnetic moments in the array of three two-layer particles. The moment configurations for the upper and lower layers are shown side by side in different colors.

*4.1. The ferromagnetic resonance in two-layer stack*

We calculated FMR spectra for different configurations of a two-layer stack. The dependencies of the FMR frequency on the distance between the layers for the stack in a state with ferromagnetic and antiferromagnetic order are presented in Fig. 9. Qualitatively, these curves are similar to the spectra of two-dimensional array of three particles (see Fig. 4), but the degree of frequency splitting for the stack is much more than for two-dimensional array.

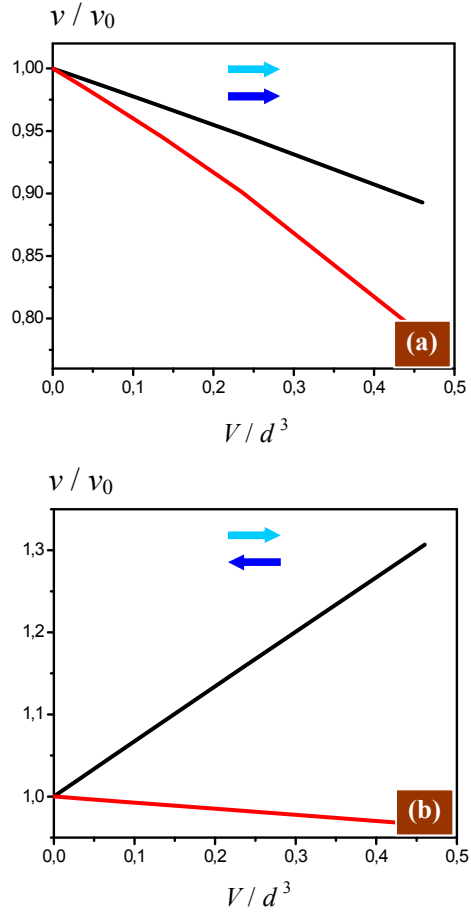

**Fig. 9.** The dependences of normalized FMR frequencies on the separation between layers in the two-layer stack for different spatial configurations of moments. (a) is the case of ferromagnetic ordering, (b) is for antiferromagnetic ordering.

*4.2. Magnetization reversal in two-layer particle array*

We have calculated the hysteresis curves and the transition schemes between different states for two-layer particle arrays for different values of the $r/d$ ratio (Fig. 10(a-c)). We show specifically results for the $\alpha=0.5$ case, for which a single layer three island cluster displayed a continuous transition between type (A) moment configurations (see Fig. 3c). The effect of increasing the ratio of the interlayer to intralayer coupling can be seen in the sequence Fig. 3c (equivalent to $r/d = 0$), Fig. 10a ($r/d = 0.45$), Fig. 10b ($r/d = 0.58$) Fig. 10c ($r/d = 0.83$). At low interlayer interaction ($r/d = 0.45$) we have the situation similar to the one-layer system (Fig. 3c). In this case there is only one stable state with quasi-uniform orientation of magnetic moments in layers and ferromagnetic interlayer ordering. The intermediate case (Fig. 10b) has two stable states, which correspond to ferromagnetic and antiferromagnetic interlayer ordering. Increasing the interlayer coupling further (Fig. 10c) leads to a closing up of the hysteresis loop, and the formation of an antiferromagnetic ordered state with zero averaged magnetic moment in zero field. Thus the variation of distance between layers enables the effective controlling the magnetic states in two-layer system.

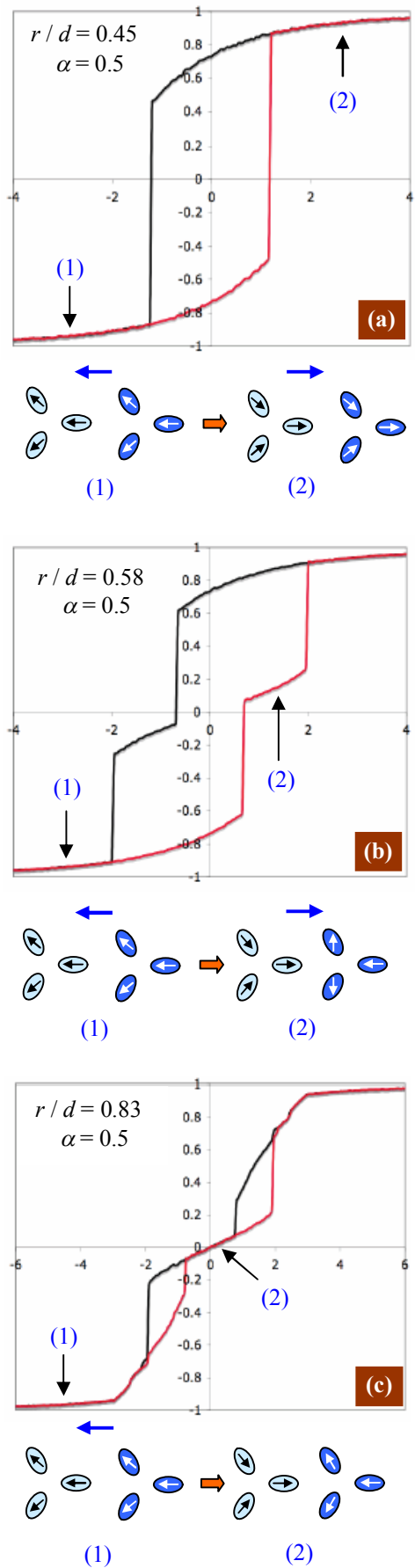

**Fig. 10.** The hysteresis curves and the schemes of transitions between different states of two-layer particle arrays for the case $\alpha = 0.5$ and two interlayer spacings: (a) $r/d = 0.45$; (b) $r/d = 0.58$; (c) $r/d = 0.83$. The moment configurations for the upper and lower layers are shown side by side in different colors.

*4.3. The ferromagnetic resonance in two-layer systems*

We have calculated the spectra for two stable states of two-layer particles array ordered on the lattice with triangular symmetry (Fig 11(a,b)). The calculations showed that the two-layer arrays have a more complicated splitting of the FMR spectrum in comparison with one-layer system. In particular the dependencies of normalized FMR frequencies of the array in the antiferromagnetic interlayer ordering have the local extremes at some distances between the layers (Fig. 11(b)).

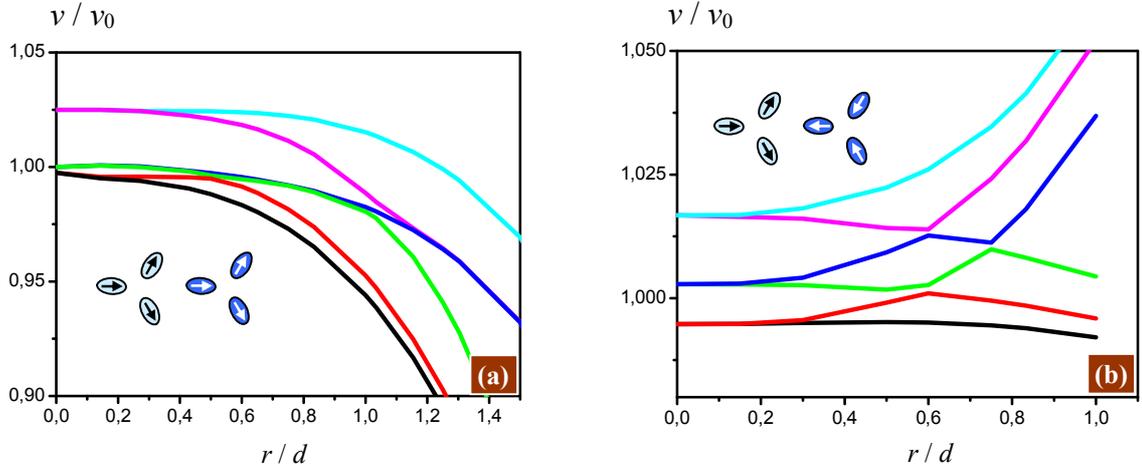

**Fig. 11.** The dependences of normalized FMR frequencies on the separation between layers in the two-layer three particle groups for different spatial configurations of moments. (a) is for ferromagnetic interlayer ordering. (b) is for antiferromagnetic interlayer ordering.

## 5. The three-layer systems

The energy of interaction in multilayer systems can be controlled also by variation of the magnetic moment magnitude for the particles located in different layers [21]. Fig. 12 shows a schematic drawing of a system of three particles arranged one above the other, and the state diagram of this system depending on the interlayer distance and the ratio of the magnetic moments of the particles. We assumed that the moments of the outer particles are the same. In such system there are three stable configurations with different magnetostatic energy. As can be seen from Fig. 12b, at the critical point with the parameters $d_{12}/d_{13} = 0.5$ and $m_2/m_1 = 0.2$ there is a frustrated state, in which the energies of all magnetic moments configurations are equal.

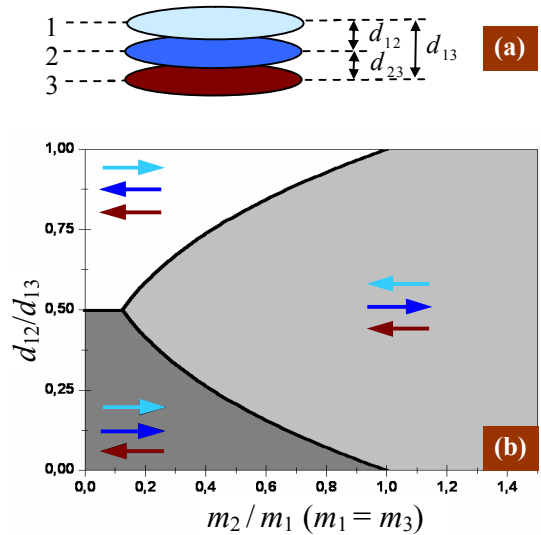

**Fig. 12.** The three-layer stack (a) and the diagram of possible states in dependence of magnetic moments relation and separations between layers (b).

5.*1. The ferromagnetic resonance in three-layer stack*

We calculated the FMR spectra for different configurations of a three-layer stack. Fig. 13 shows the dependences of the FMR frequency on the distance between the layers for the stack in a state with ferromagnetic, antiferromagnetic and mixed ferromagnetic - antiferromagnetic ordering. Qualitatively, these curves are similar to the spectra of the two-dimensional array of three particles (Fig. 4), but the degree of frequency splitting for the stack is much more than for two-dimensional array.

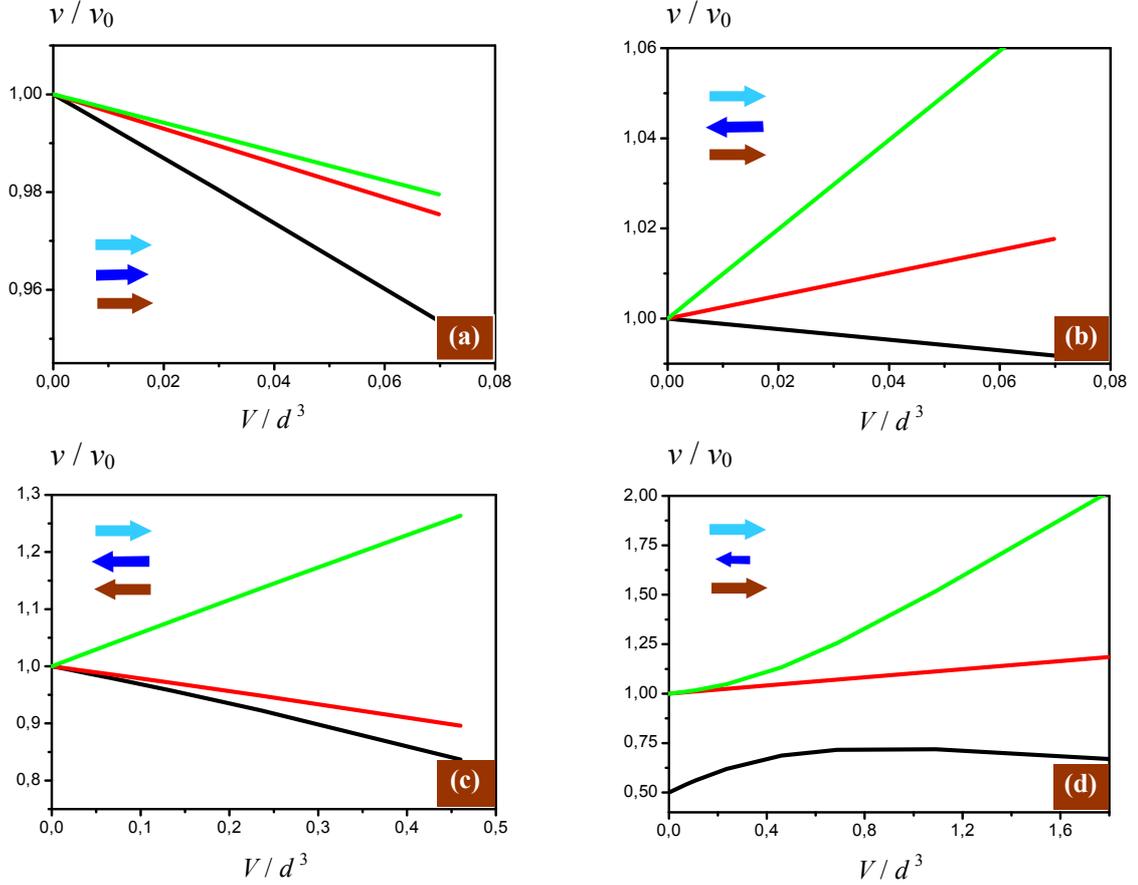

**Fig. 13.** The dependences of normalized FMR frequencies on the separation between layers in the three-layer stack for different spatial configurations of moments. (a) is the case of ferromagnetic ordering, (b) is for antiferromagnetic ordering, (c) is for mixed ferromagnetic - antiferromagnetic ordering and (d) is for the stack with the reduced magnetic moment of the middle layer ($m_1 = m_3$, $m_2/m_1 = 0.5$).

Also we have considered the stack with different magnetic moments in the layers (Fig. 13d). The relation of magnetic moment of central layer $m_2$ to moments of outer layers ($m_1 = m_3$) is equal 0.5. In this case we have initial splitting due to the difference of magnetic moments and the dependences of normalized FMR frequencies have a local extremum. Thus allowing differences between the magnetic moments of layers provides an additional way of controlling the FMR spectrum.

*5.2. Magnetization reversal in three-layer particle array*

The results of hysteresis calculations for three-layer stacks of three identical particle arrays on a triangular lattice are presented in Fig. 14. We again consider the α=0.5 case and display the hysteresis plots for different interlayer spacing: (a) $r/d = 0.45$; (b) $r/d = 0.58$; (c) $r/d = 0.83$. When the interlayer coupling is not strong ($r/d = 0.45$), the system implements a simple magnetization reversal process through the state with antiferromagnetic ordering of the magnetic moments in neighboring layers (Fig. 14a). Increasing the interlayer interaction leads to additional configuration states encountered during the hysteresis cycle, as shown in Fig. 14b for $r/d = 0.58$. In this case the state (1) becomes unstable. The transition from state (1) to state (2) involves just a flip of the moment of one particle in the central layer (indicated in red in scheme). The other two particles in the central layer then experience a flip in the transition to state (3). The low average moment state (3) is essentially the same as state (2) in Fig. 14a. The small average moments in the two plots are slightly different because of slight differences in deviations from easy axis alignment. The transition from state (3) to state (4) involves moment reversals in all three layers, so we have the same state as (3) but with the opposite directed average magnetic moment. Fig. 14c shows the behavior as the interlayer coupling is increased still further to $r/d = 0.83$. The transition from state (1) to state (2) is similar to that in Fig. 14b, but in this case state (2) becomes unstable. Here we have only one antiferromagnetic ordered stable state (3) similar to the state (3) in Fig. 14b. Again the transition from state (3) to state (4) involves moment reversals in all three layers, so we have practically the same state with the opposite directed average magnetic moment.

As the number of layers is increased, the number of states traversed in a hysteresis cycle also increases, reflecting the more complex sequence of moment reorientation processes that is taking place.

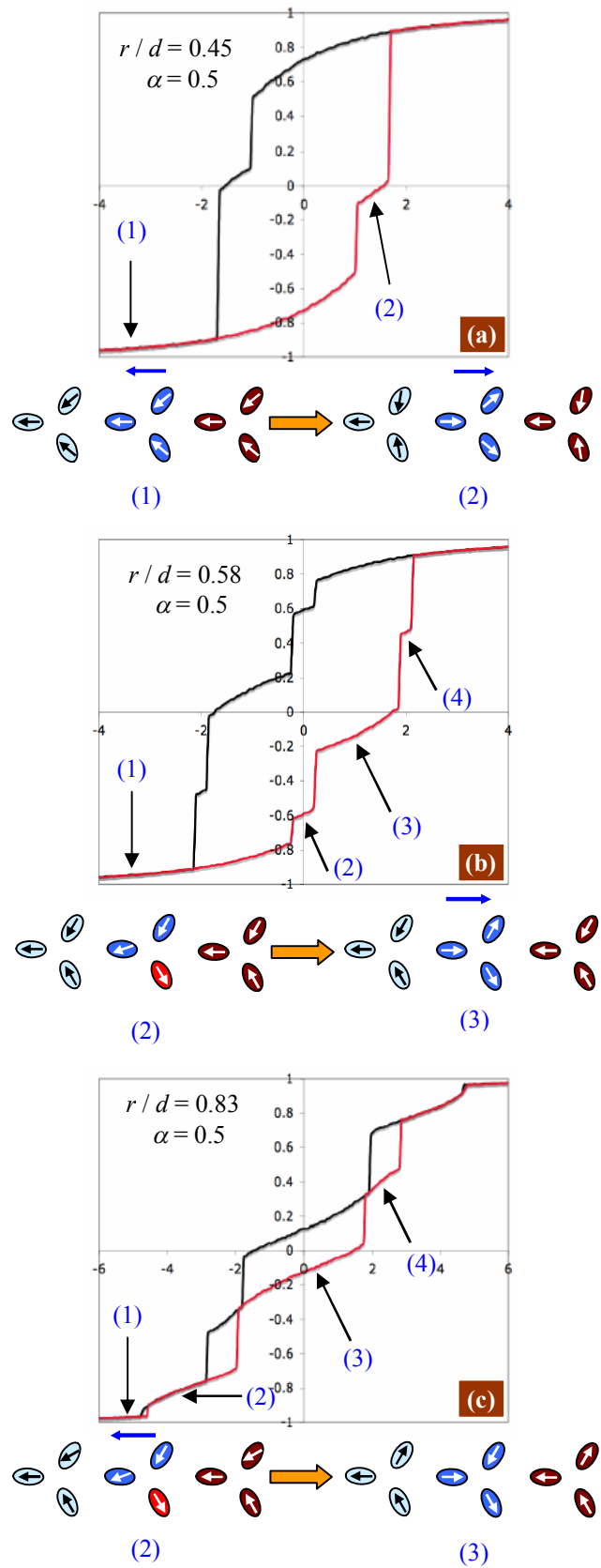

**Fig. 14.** The hysteresis curves and the schemes of transitions between different states of three-layer particles array for $\alpha = 0.5$ case: (a) $r/d = 0.45$; (b) $r/d = 0.58$; (c) $r/d = 0.83$. The configuration of the magnetic moments of the particles corresponds to the points on the hysteresis curve. The one-particle flip in transition schemes in the pictures (b) and (c) is indicated in red.

*5.3. The ferromagnetic resonance in three-layer system*

The Fig. 15 shows the dependencies of the FMR frequencies on the distance between the layers for three-layer particle array in the state with ferromagnetic (Fig. 15a), antiferromagnetic interlayer order (Fig. 15b), in the state with partial ferromagnetic and antiferromagnetic ordering (Fig. 15c) and for the system with different magnetic moments in antiferromagnetic configuration (Fig. 15d).

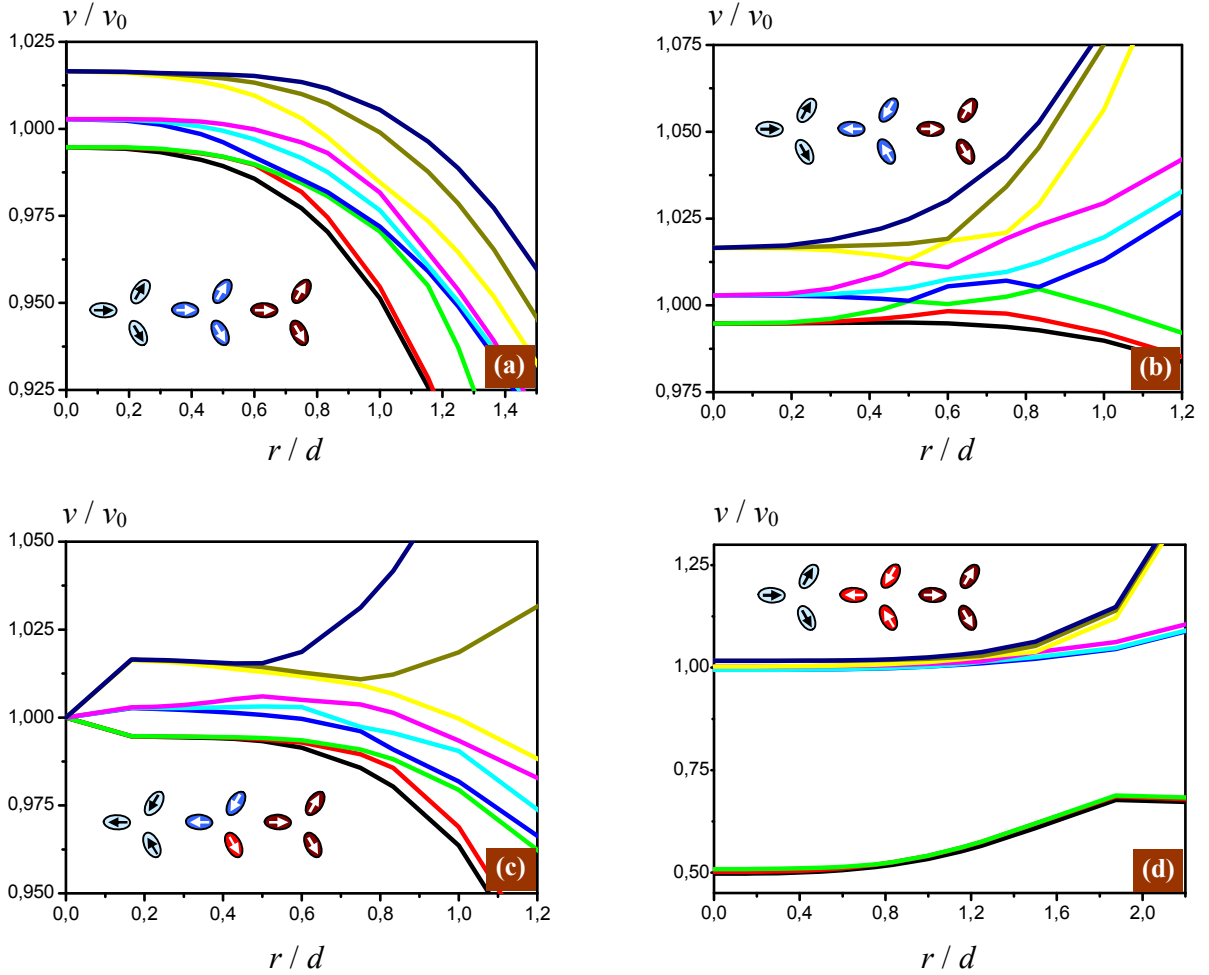

**Fig. 15.** The dependences of normalized FMR frequencies on the separation between layers in array of the three-layer stacks for different spatial configurations of moments. (a) is for the ferromagnetic ordering. (b) is for antiferromagnetic ordering. (c) is for the state with partial ferromagnetic and antiferromagnetic ordering. (d) is for the stacks with the reduced magnetic moment of the middle layer ($m_1 = m_3$, $m_2 / m_1 = 0.5$) in the state with antiferromagnetic interlayer ordering. The middle layer with the reduced magnetic moment is highlighted in red.

Analogous to the two-layer system the dependencies of FMR spectra on the distance between the layers for three-layer system are nonmonotonic as a consequence of competition between the interlayer and interlayer interactions. Note that the use of the stacks with different layer thickness or different layer materials substantially extends the possibilities in controlling of ferromagnetic resonance in such systems.

**6. Discussion**

Thus, the calculations showed that the magnetostatic interaction between the particles in the layer and the interlayer interaction strongly affect the magnetization reversal dynamics in an external magnetic field and lead to the complex splitting of FMR spectra.

In the calculations, we used a simple model of anisotropic point dipoles. However, the real particles are distributed systems with a complex character of the magnetic interactions and this leads to inhomogeneous magnetization reversal modes and complication of the FMR spectrum. Along with the uniform coherent precession modes in such systems, there exist nonhomogeneous modes [22-24] associated with the internal spin resonances and edge effects. However for the systems consisting of small magnetic nanoparticles the FMR spectrum of ground modes of uniform precession and the basic mechanisms of the splitting of resonance frequencies are described adequately by the simple model we have used.

It was shown, that the systems consisting of one-layer elements with only in-plane interactions display the complicated but weak splitting of FMR spectra (Figs. 4,6). The arrangement of elliptical discs in multilayer stacks systems leads to the considerable increase of interparticle magnetostatic interaction and as a consequence to the appearance of unusual states during magnetization reversal and significant changes in the FMR spectra. It allows us to construct the different systems with fine tuning of ferromagnetic resonance. For example, we have demonstrated that in two-layer system (consisting of three two-layer stacks, located at 120°) there are two stable states (Fig. 10). The first state has a quasi-uniform moment's distribution (with ferromagnetic interlayer ordering) and can be realized in a system with low interlayer interaction (Fig. 10a). The second state has zero average magnetic moment (with antiferromagnetic interlayer ordering) and can be realized in the system with high interlayer interaction (Fig. 10c). In the system with intermediate interlayer interaction there are both these states, which can be switched by magnetization reversal in an external magnetic field (Fig. 10b). The FMR spectra corresponding to these two states are presented in Fig. 11 and have the fundamental differences. The quasi-uniform state has the spectrum shifted to lower frequencies, while the state with zero average moment has the spectrum with four branches shifted to higher frequencies. Therefore such two-layer particles arrays with simply recognized magnetic states can be used in the magnetic labeling systems.

We have shown that adding a third layer in the stack leads to an extra frustration of magnetic states and complicated splitting in FMR spectra. Depending on the interlayer coupling in such systems the states with ferromagnetic, antiferromagnetic and mixed ferromagnetic - antiferromagnetic interlayer ordering are realized (Fig. 14). This significantly expands the possibilities to control of the FMR spectrum in such systems (Fig. 15(a-c)). Moreover the FMR spectrum splitting in three-layer systems is considerably larger than in the case of two-layer systems.

The great perspectives for the engineering of magnetic states and FMR spectra are associated with the possibility of varying the ratio of magnetic moments for the particles belonged to different layers and interlayer spacings. In particular, we have considered the three-layer system with a reduced magnetic moment of the particles in the middle layer. Such systems can achieve an increase the splitting of the spectra due to the initial frequency splitting for the differed particles forming the array (Fig. 15d).

## 7. Conclusion

Thus we have demonstrated that multilayer ferromagnetic nanoparticles significantly expand the opportunities to create the magnetically frustrated systems. The different spatial arrangement, variation of magnetic moments and interlayer spacings enable the fine tuning of possible stable magnetic states and effective control of FMR spectrum in such systems that is very important from the standpoint of practical applications.

**Acknowledgements**

Authors thanks K.D. Bessmertniy for assistance and A.A. Fraerman for the useful discussions.
This work was supported by grants of Russian Foundation for Basic Research, the programs of Russian Academy of Sciences and The Ministry of Education and Science of Russian Federation.